\documentclass[aps,pre,twocolumn,showpacs,groupedaddress,superscriptaddress,longbibliography]{revtex4-1}

\usepackage[english]{babel}
\usepackage[utf8]{inputenc}
\usepackage{graphicx}
\usepackage[caption=false]{subfig}
\usepackage{amsmath}
\usepackage{amsfonts}
\usepackage{amssymb}
\usepackage{color,soul}
\setulcolor{red}

\begin{document}


\title{Freezing, accelerating and slowing directed currents in real time with superimposed driven lattices}

\author{Aritra K. Mukhopadhyay}
 \email{Aritra.Mukhopadhyay@physnet.uni-hamburg.de}
 \affiliation{Zentrum f\"ur Optische Quantentechnologien, Universit\"at Hamburg, Luruper Chaussee 149, 22761 Hamburg, Germany}
 \author{Benno Liebchen}
  \email{Benno.Liebchen@ed.ac.uk}
 \affiliation{SUPA, School of Physics and Astronomy, University of Edinburgh, Edinburgh EH9 3FD, United Kingdom} 
 \author{Thomas Wulf}
 \email{Thomas.Wulf@physnet.uni-hamburg.de}
 \affiliation{Zentrum f\"ur Optische Quantentechnologien, Universit\"at Hamburg, Luruper Chaussee 149, 22761 Hamburg, Germany}
 \author{Peter Schmelcher}
  \email{Peter.Schmelcher@physnet.uni-hamburg.de}
 \affiliation{Zentrum f\"ur Optische Quantentechnologien, Universit\"at Hamburg, Luruper Chaussee 149, 22761 Hamburg, Germany}
 \affiliation{The Hamburg Centre for Ultrafast Imaging, Universit\"at Hamburg, Luruper Chaussee 149, 22761 Hamburg, Germany}


\date{\today}

\begin{abstract}
We provide a generic scheme offering real time control of directed particle transport in superimposed driven lattices. This scheme allows to accelerate, slow and freeze the transport on demand, by switching one of the lattices subsequently on and off. The underlying physical mechanism hinges on a systematic opening and closing of channels between transporting and non-transporting phase space structures upon switching, and exploits cantori structures which generate memory effects in the population of these structures. Our results should allow for real time control of cold thermal atomic ensembles in optical lattices, but might also be useful as a design principle for targeted delivery of molecules or colloids in optical devices. 
\end{abstract}

\pacs{05.45.Gg, 05.60.Cd, 05.45.Pq, 05.45.Ac}

\maketitle

\section{Introduction}
Temporally driven lattice potentials have attracted considerable attention in recent years \cite{Salger2009,Denisov2006,Brown2008,Gommers2008,Reimann2002,Hanggi2009,Dittrich2015,Cubero2016} as their experimental controllability allows for an insightful approach into the complex world of non-equilibrium physics.
A phenomenon of particular interest in these systems 
is the ratchet effect. 
Here, the breaking of certain spatio-temporal 
symmetries of the system allows one to convert unbiased fluctuations into directed particle motion 
even in the absence of mean forces \cite{Denisov2014,Denisov2008,Flach2000,Gommers2005}.
This can be seen as a working principle of a motor operating on smallest scales relevant to phenomena ranging from 
intracellular transport problems \cite{Julicher1997} 
and cancer cell metastasis \cite{Mahmud2009} to
the transport of colloidal particles \cite{Faucheux1995,Rousselet1994} in optical lattices or vortices in 
Josephson junction arrays \cite{Majer2003}.
Novel ratchet experiments using atomic ensembles in ac-driven optical lattices 
\cite{Schiavoni2003,Lebedev2009} allow for an admirable controllability
both in the ultracold quantum regime \cite{Salger2009} and at 
micro kelvin temperatures where a classical dynamics approach successfully describes experiments \cite{Brown2008,Renzoni2009}.
Naturally in view of their widespread applications, the controllability of directed particle currents 
has been a focal point of research since the early days of ratchet physics. 
Here, owing to the absence of an obvious force bias even the transport direction is sometimes difficult to predict
and numerous cases of `current reversals' have been reported where the direction of the transport in the asymptotic time limit could be 
reversed by changing a control parameter even though the symmetries of the system remain unaffected \cite{Dandogbessi2015,Arzola2011,Denisov2007,Schreier1998,DeSouzaSilva2006,Marconi2007,Spiechowicz2015,Cubero2012,Liebchen2012}.
A limitation of most of these schemes is that only the asymptotic transport direction can be controlled
rather than allowing for real time control of the current which would be certainly desirable in order
to apply ratchets as nanomotors \cite{Hanggi2009} and to problems like targeted drug delivery \cite{Hoffmann2012}. A recent exception is \cite{Arzola2011} which requires, however, dissipation and is restricted to `flipping' the directed current at fixed strength.

\begin{figure}
\includegraphics[scale=0.06]{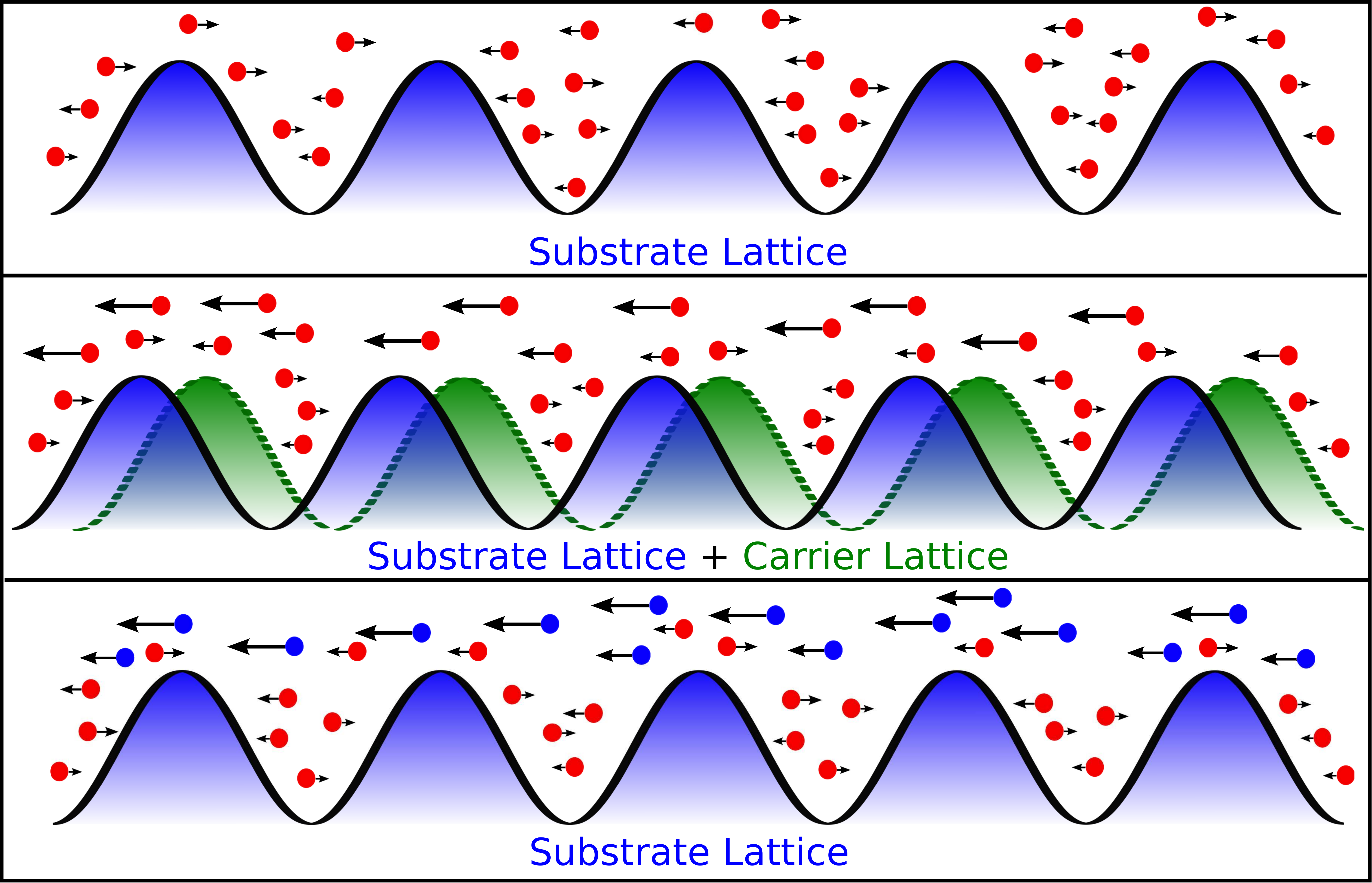}
\caption{\small Schematic diagram of the setup and real time control of directed transport. 
Upper panel: Non-transporting state in the oscillating substrate lattice. Middle panel: Directed transport after switching on the carrier lattice. 
Lower panel: 
Persistent transport after switching off the carrier lattice. 
Red particles perform diffusive motion whereas the blue ones are ballistic. 
The length and direction of the arrow indicate the speed and direction of the particle respectively.}
\label{fig1}
\end{figure}

Here, we exemplify a generic route towards the real-time control of directed currents. This allows not only to dynamically control both direction and strength of the transport, up to unusually high efficiencies, but also to freeze the transport velocity on demand.  
Using one non-transporting and symmetric oscillating lattice as a `substrate' for particles (Fig.~\ref{fig1}, upper panel), we subsequently switch a second oscillating lattice, 
called the `carrier' lattice, 
on and off. In particular, switching the carrier lattice on breaks the parity and time-reversal symmetries of our setup and induces a directed particle current (middle panel) 
accelerating the transport in a direction which can be
controlled by the phase difference between the carrier and the substrate lattice. 
Switching the carrier lattice off does not lead to a decay of the transport towards zero but `freezes' it at constant velocity (lower panel).
This can be repeated many times and allows one to design transport in real time. 
As the underlying mechanism, we identify 
a systematic opening and closing of cantorus structures, acting as barriers between transporting and non-transporting phase space structures upon switching.
Thereby, the timescale on which the current can be manipulated is set by the flux through the cantorus and we show that manipulations are, in fact, possible for up to $\sim 10^5$
driving periods.
%
Our scheme does not require noise, but is robust to it, and is designed for straightforward implementation with cold thermal atoms in 
superimposed driven optical lattices where state of the art technologies allow to avoid 
interference terms between both lattices. Here, our scheme can be applied to guide atomic ensembles through optical lattices
on paths which can be designed in real time. The underlying working principle should be of more general relevance, for example, 
as a design principle for real time controlled targeted delivery of molecules or colloids in optical lattices or, possibly, also on other vibrated substrates.

\section{Setup} 
We consider non-interacting classical particles of mass $m$, position $x$ and momentum $p$, 
described by a single particle Hamiltonian $H(x,p,t)=\frac{p^2}{2m}+V(x,t)$,
in a periodic potential $V(x,t)=V_S(x,t) + V_C(x,t)$. Here, $V_S$ represents the `substrate lattice' and $V_C$ the `carrier lattice' with
\begin{eqnarray}
V_S(x,t)&=& V_S\cos^{2}[k(x+d \cos(\omega t))] \label{pot} \\
V_C(x,t)&=& V_C\cos^{2}[k(x+2d \cos(2\omega t+\phi)) +\delta] \nonumber.
 \end{eqnarray}
Both lattices have identical wavenumber $k$, but the oscillation amplitude $d$ and frequency $\omega$ of the carrier lattice are twice as large as those for the substrate, which leads
to spatial and temporal periodicities of $L=2\pi / k$ and $T=2\pi / \omega$ of $H$.
Clearly, after averaging over time and space, this system is force free and hence unbiased.
Our Hamiltonian may describe, for example, cold atoms in the classical regime of $\mu K$ temperatures \cite{Brown2008,Renzoni2009}
exposed to two counterpropagating laser beams of perpendicular polarization, preventing the occurrence of interference terms
in Eq.~(\ref{pot}). The lateral oscillation of both lattices can be achieved by phase modulating both 
laser beams using standard techniques like acousto-optical modulators and radio frequency generators (see e.g. \cite{Schiavoni2003,Wickenbrock2012}).
\\To identify the relevant control parameters we introduce
dimensionless variables $x'=2k x$ and $t'=\omega t$. Using
$\mu= \frac{m\omega^{2}}{2V_{S}k^{2}}$, $\nu=2k d$, 
$\gamma_V=\frac{V_C}{V_S}$, we get the equation of motion
\begin{equation}\label{eq2} 
\mu\ddot{x}= \sin (x+\nu \cos t) + \gamma_V \sin (x+2\nu \cos(2t+\phi) +\delta)
\end{equation}
where we omitted the primes on $t'$ and $x'$.
\begin{figure}
\includegraphics[scale=0.088]{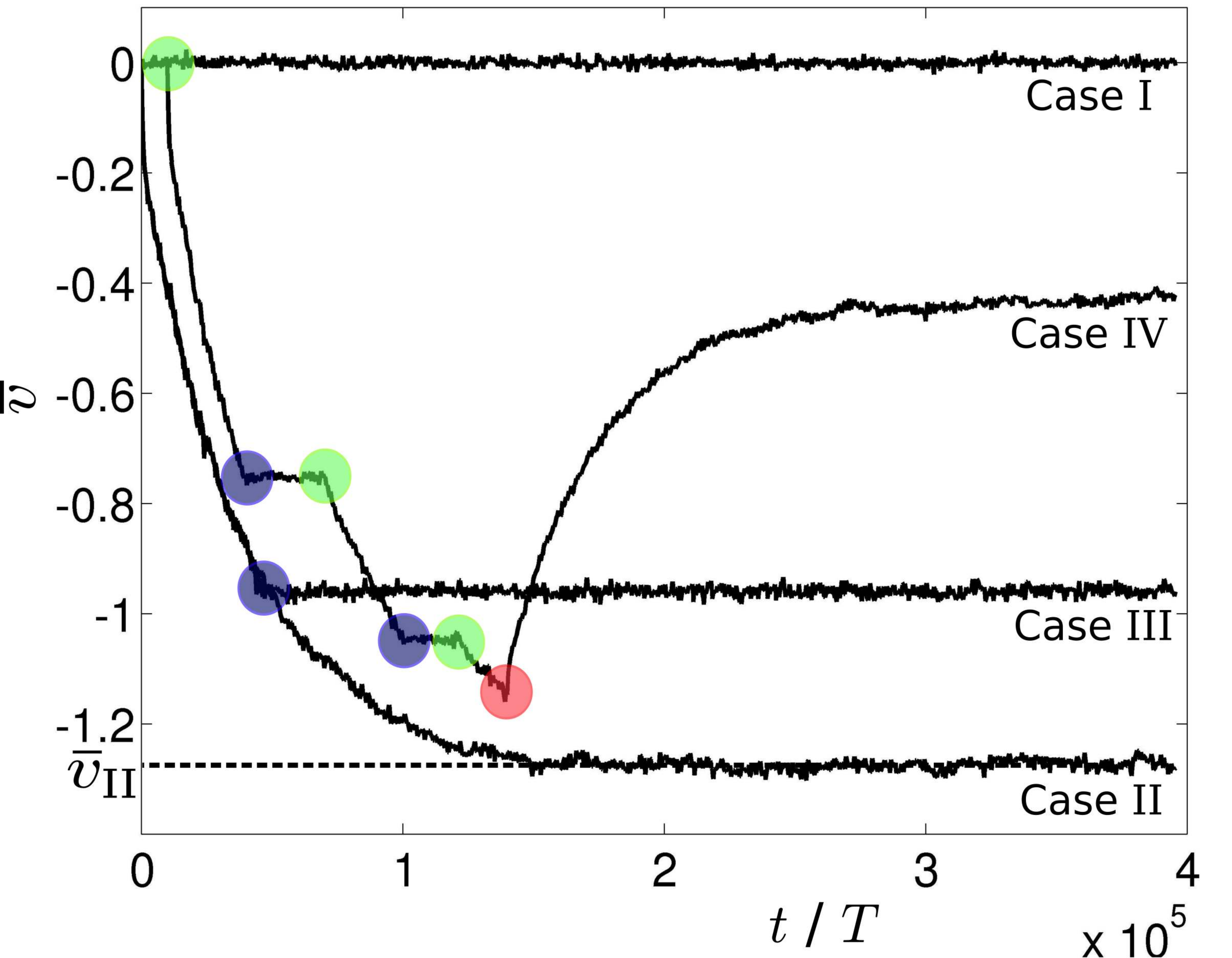}
\caption{\small Mean transport velocity $\bar v$ of a particle ensemble as a function of time for four different cases. (I): Only substrate lattice ($\gamma_V=0$). 
(II): Substrate and superimposed carrier lattice $\gamma_V=1$. (III): Both lattices but carrier lattice switched off at $t=0.11 t_{\rm tot}$  (blue dot)  ($\gamma_V=1$).
(IV): Subsequent switches of the carrier lattice; blue dots show times ($t=0.10 t_{\rm tot}$ and $t=0.25 t_{\rm tot}$) where the carrier lattice is switched off, green dots when it is switched on ($t=0.025 t_{\rm tot}$, $t=0.175 t_{\rm tot}$ and $t=0.30 t_{\rm tot}$). 
At the final switch (red dot; $t=0.35 t_{\rm tot}$), we switched also the relative driving phase $\phi$ from $\phi=\pi/2$ to $\phi=-\pi/2$.
Remaining parameters: $\mu=1.2665$, $\nu=\pi$, $\delta=\pi/2$ and $\phi=\pi/2$. }
\label{fig2}
\end{figure}

\section{Ratchet transport and lattice switches} 
In order to explore the transport properties of our setup, we propagate $N=2\times 10^{4}$ particles up to $t_{\rm tot}=4\times 10^{5}T$ by numerical integration 
of Eq.~(\ref{eq2}) using a Runge-Kutta Dormand Prince integrator \cite{Dormand1980}. The initial velocities of the particles are chosen randomly within the low velocity regime such that their initial kinetic energies are small compared to the potential height of both lattices. In this section, we present the main results and discuss the underlying physical mechanisms in the following sections. 

In the case of only a substrate lattice being present, we do not observe directed transport (case I in Fig.~\ref{fig2}). This is to be expected, because the corresponding equation of motion (Eq.~(\ref{eq2}) with $\gamma_V=0$) is invariant under time reversal: $t\rightarrow -t$, thus preventing directed particle motion in unbiased systems \cite{Flach2000}. Applying the carrier lattice additionally ($\gamma_V=1$ in Eq.~(\ref{eq2})) and choosing appropriate `phase shifts' to the substrate ($\phi \neq 0, \pi$ and $\delta \neq 0, \pi$) allows one to break both time reversal and parity symmetry which leads indeed to directed transport (case II in Fig.~\ref{fig2}). This transport slowly accelerates and finally saturates at $\bar v_{\text{II}}\simeq -1.25$ which is comparable to the spread of the velocity distribution of the particles. This constitutes an unusually high efficiency for a Hamiltonian ratchet, where the mean drift velocity is typically one or two orders of magnitude less than the standard deviation of the particle velocity distribution. We now consider the same situation, but switch off the carrier lattice, instantaneously, at $t=0.11 t_{\rm tot}$ (blue dot; case III in Fig.~\ref{fig2})). Interestingly, after switching off the carrier lattice the transport persists. Rather than decaying back towards zero, as one might expect for a symmetric setup, it does not decay but is frozen at its value of the time of the switch. That is, our atomic ensemble travels with constant average speed through the symmetrically oscillating lattice. Remarkably, this allows for an intriguingly simple real time control of the transport velocity: Once the desired transport is achieved one simply needs to switch off the carrier lattice.

We now consider a similar case, but switch the carrier lattice subsequently on and off (case IV in Fig.~\ref{fig2}). Once the transport has been frozen for a while we can accelerate it by switching on the carrier lattice again (second green dot on case IV curve). Switching it off, for a second time (second blue dot), freezes also this enhanced transport at a constant strength. Clearly, to achieve a highly flexible real time control of the transport it would be desirable to be able to slow it down. This can in fact be achieved by switching on the carrier lattice again, but this time with a phase difference of $\phi=-\pi/2$ to the substrate (red dot). 
We see in Fig.~\ref{fig2} that this indeed slows the transport systematically down. Overall, we demonstrated a remarkably simple protocol allowing to enhance, freeze or slowdown the transport of atomic ensembles in two optical lattices on demand. 

\begin{figure}
\includegraphics[scale=0.072]{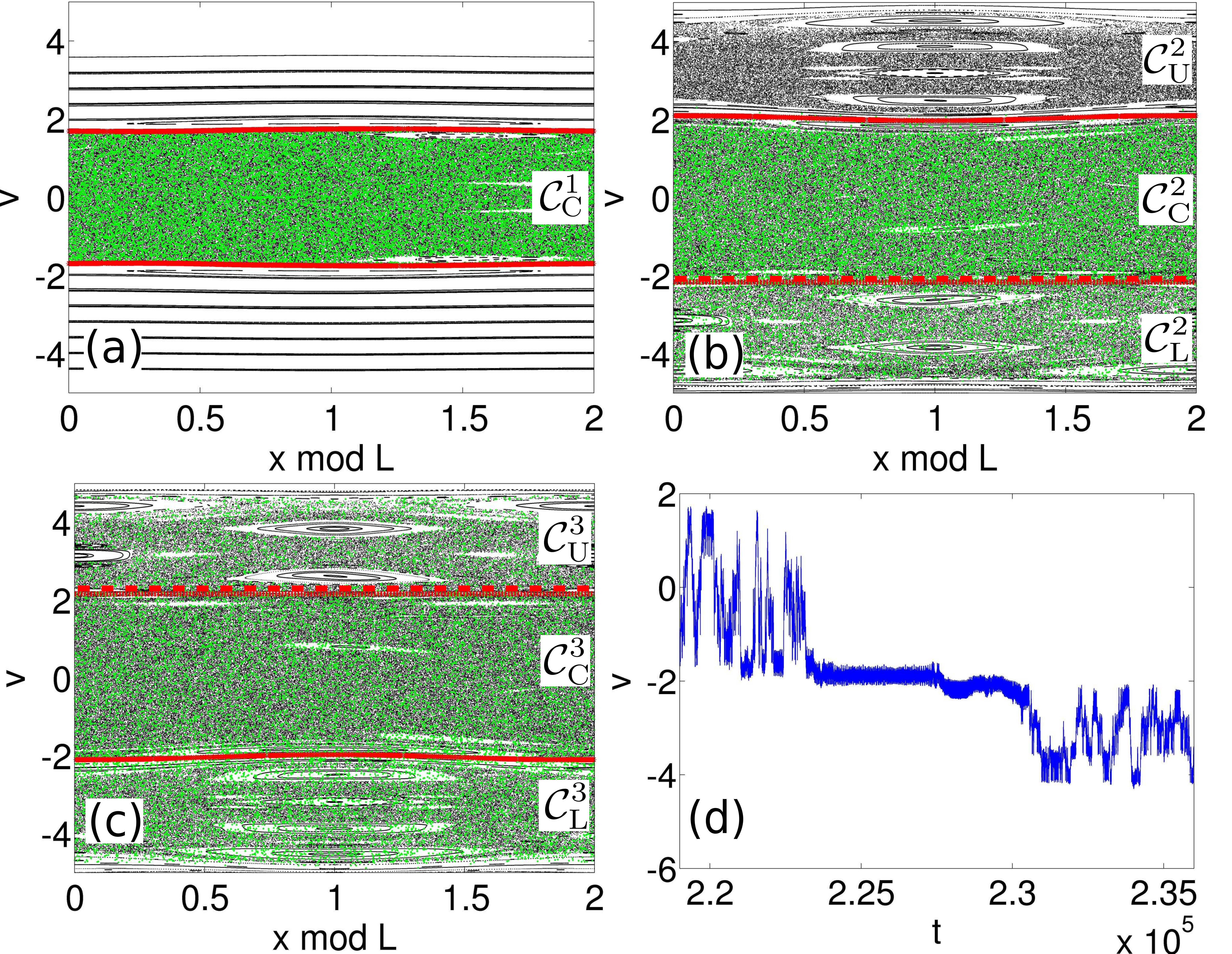}
\caption{\small The position (mod L) and velocity of all the $N$ particles (green) at (a) $t=0.04 t_{\rm tot}$ in case I superposed on the PSOS P1(black dots and lines) of the substrate lattice (b) $t=0.42 t_{\rm tot}$ in case III superposed on the PSOS P2(black dots) corresponding to both the substrate and the carrier lattices and (c) $t=0.95 t_{\rm tot}$ in case IV superposed on the PSOS P3(black dots) corresponding to both the lattices but with  $\phi=-\frac{\pi}{2}$. Red solid lines denote the position of the FISCs whereas the red dashed lines indicate the location of the cantorus (see text). $\mathcal{C}^i_{\rm U,C,L}$, $i=1,2,3$ denotes the upper, central and lower chaotic layer of P1, P2, and P3 respectively. (d) A zoom into the typical trajectory of a particle initiated at low velocity in the central chaotic sea of the PSOS P2 in Fig.~\ref{fig3}b, showing the particle's stickiness to the cantorus.}
\label{fig3}
\end{figure}

\section{Discussions}
\subsection{Phase space analysis}
\label{phase space analysis}
It turns out that the physical mechanism underlying the real time control of directed currents we just demonstrated crucially hinges on the mixed phase space structure of our two lattice system. Hence to understand it we perform a systematic analysis of its microscopic composition and analyze its dynamical occupation by the considered particle ensemble. First, to understand the structure of the phase space itself, we take `stroboscopic' snapshots of particles with different initial conditions leading to Poincar\'{e} surfaces of sections (PSOS) which provide a representative overview of the structure of the complete 3D phase space \cite{Tabor1989}. We also exploit the spatial periodicity of our setup and project the particle position back to the first unit cell $x \in [0,L)$ of the lattice. The PSOS of the substrate lattice (henceforth referred to as P1) is symmetric about $v=0$ (Fig.~\ref{fig3}a black dots) and contains a large central `chaotic sea' $\mathcal{C}^1_{\rm C}$ between $v\simeq \pm 1.6$. On top of the PSOS, we show the snapshot of the particle coordinates, at a given time, used to determine $\bar{v}$ in case I of Fig.~\ref{fig2} (green dots in Fig.~\ref{fig3}a) illustrating the uniform symmetric chaotic diffusion of particles through the lattice resulting in no transport. The chaotic sea is bounded by the first invariant spanning curves (FISC; red lines in Fig.~\ref{fig3}a) which prevents acceleration of our low velocity initial conditions beyond $|v| > 1.6$. Contrarily, a particle with initial condition on one of the regular invariant curves at $|v| \gtrsim 1.6$ (black lines in Fig.~\ref{fig3}a) shows ballistic unidirectional motion through the lattice. 

Let us now explore how the phase space structure changes in presence of the carrier lattice. Most prominently, the two lattice PSOS (black dots in Fig.~\ref{fig3}b), henceforth referred to as P2, is not mirror symmetric about the $v=0$ axis. Besides the chaotic sea $\mathcal{C}^2_{\rm C}$ at small velocities, it exhibits two additional chaotic layers at higher velocities: the upper layer $\mathcal{C}^2_{\rm U}$ at $v \gtrsim 2.2$ and the lower layer $\mathcal{C}^2_{\rm L}$ at $v \lesssim -2.2$. The crucial point now is that the choice of an appropriate value for $\gamma_V$ allows one to connect $\mathcal{C}^2_{\rm C}$, asymmetrically, only with $\mathcal{C}^2_{\rm L}$ through a `cantorus' structure (red dashed line in Fig.~\ref{fig3}b) which is a hierarchical chain of stable and unstable fixed points, while it remains separated from $\mathcal{C}^2_{\rm U}$ by a regular invariant curve (red solid line in Fig.~\ref{fig3}b). This allows particles to enter $\mathcal{C}^2_{\rm L}$ but not $\mathcal{C}^2_{\rm U}$. This can be easily seen from the fixed time snapshot of the particle distribution onto P2 corresponding to case III denoted by the green dots in Fig.~\ref{fig3}b. These particles in  $\mathcal{C}^2_{\rm L}$ still move irregularly but now only in one direction through the lattice, which is the origin of the transport we observed in Fig.~\ref{fig2}. The fact that the transport velocity does not quickly converge to a constant velocity, but accelerates very slowly, on time scales of $10^5$ driving periods towards its asymptotic value (case II in Fig.~\ref{fig2}) is owed to the cantorus linking $\mathcal{C}^2_{\rm C}$ and $\mathcal{C}^2_{\rm L}$, which effectively acts as a semi-permeable barrier to the particles approaching it and slows down the uniform filling of the accessible parts of the phase space. 

\subsection{Conversion between diffusive and ballistic motion} 
To understand how switching the carrier lattice subsequently on and off allows to freeze, accelerate and revert the directed transport, we now analyse the 
impact of lattice switches on the population of phase space structures (green dots in Fig.~\ref{fig3}a-c).

In Fig.~\ref{fig2}, case III, when we froze the directed transport by switching off the carrier lattice, the phase space changed suddenly from P2 to P1.
The crucial observation is now that particles located in $\mathcal{C}^2_{\rm L}$ of P2 at the instance of the lattice switch (green dots below $v\sim -1.6$ in Fig.~\ref{fig3}b) are located in the regular domain of spanning curves in P1 (see Fig.~\ref{fig3}a) after the switch. As usual, particles which are located on regular spanning curves after the switch (black lines in Fig.~\ref{fig3}a), are confined to these structures and travel with almost constant velocity through the lattice. Hence, also the ensemble averaged velocity, i.e. the directed transport, remains approximately constant or `frozen' which explains our observation in Fig.~\ref{fig2}, case III (the particles in the chaotic sea $\mathcal{C}^1_{\rm C}$ of P1 do not contribute to the transport as P1 is symmetric around $v=0$). In conclusion, the instantaneous switch of the dynamical system has caused a conversion from diffusive to regular motion for some particles, which is reminiscent to the conversion processes between regular and ballistic dynamics observed in driven superlattices \cite{Wulf2012,Petri2011}.

It is now straightforward to see how switching the carrier lattice for a second time (case IV in Fig.~\ref{fig2}) accelerates the transport again. This switch suddenly changes the underlying phase space from P1 to P2 and connects, again, $\mathcal{C}^2_{\rm C}$ with $\mathcal{C}^2_{\rm L}$. Hence, since the particle density in $\mathcal{C}^2_{\rm C}$ is (still) higher than the density in $\mathcal{C}^2_{\rm L}$ (Fig.~\ref{fig3}b), particles continue penetrating through the cantorus into $\mathcal{C}^2_{\rm L}$ which stops only for a uniform particle distribution over the phase space. Furthermore, there is now a natural way to slow down the transport. Choosing an inverse phase difference of $\phi=-\pi/2$ (instead of $+\pi/2$), mirrors P2 around $v=0$ and the particles now slowly redistribute from the central chaotic layer $\mathcal{C}^3_{\rm C}$ into the upper chaotic layer $\mathcal{C}^3_{\rm U}$ as shown in the PSOS P3 (black dots in Fig.~\ref{fig3}c). This creates a `counterweight' to the particles in $\mathcal{C}^3_{\rm L}$ which slowly, but continuously grows (for a snapshot see green dots in Fig.~\ref{fig3}c), which explains the observed decrease of the directed transport. 

How long can we proceed to accelerate, slowdown and revert the transport?  The timescale is set by the uniform filling of the entire chaotic sea of P2. This limiting timescale depends crucially on the flux through the cantorus which in turn can be tuned by  varying the relative strengths of both lattices i.e. by changing $\gamma_V$. For $\gamma_V=1$ (the value we used), at about $t \sim 4\times 10^5T$ the entire chaotic sea of the two lattice setup (lower, central and upper sea) is uniformly filled with particles. At this point, no further modulation of the transport is possible within our scheme.

\section{Experimental realizations}
We believe that our dynamical control of directed currents can be realized in experimental setups using cold atoms in driven optical lattices where the periodic potential is generated by counterpropagating laser beams of perpendicular polarization \cite{Schiavoni2003,Gommers2005,Wickenbrock2012,Renzoni2009}. The resulting lattice can be driven by phase modulation using acousto-optical modulators and radio frequency generators which also allow to keep both lattices in phase and to implement a driving amplitude on length scales of the order of $L$ \cite{Gommers2005,Renzoni2009}. Translating our parameters to experimentally relevant quantities for rubidium atoms, we obtain $V_S =V_C \sim 20 E_r$, $\omega \sim 10\omega_r$ and the product $dk\sim \frac{\pi}{2}$, where $E_r$ and $\omega_r$ are the recoil energy and recoil frequency of the atom respectively. These experiments operate in the demonstrated classical regime of $\mu K$ temperature {\cite{Renzoni2009}}. Even for colder temperatures, in the semiclassical regime, we expect tunnelling through cantori {\cite{Reimann2002}} which should not alter our control scheme in general but only reduce the operational timescale. We note, that our scheme can be `refreshed' by employing Sisyphus cooling, which can be used
to localize our particle ensemble in the central chaotic sea again. Notably in these experiments many particle effects are not important, but one can in principle tune parameters to probe the impact of weak interaction effects {\cite{Chin2010}}. This can have important consequences for the transport {\cite{Liebchen2015}}, but it affects the particle distribution in phase space only on long timescales {\cite{Liebchen2015}} and should therefore leave our scheme unaffected. In contrast to Brownian ratchets, our mechanism does not depend on noise and we explicitly checked that it is robust to noise of strengths in the regime typical for cold atom ratchet experiments \cite{Cubero2010}. Stronger noise would enhance the particle flux through the cantori and other regular structures significantly decreasing both the maximally achievable transport velocity and the operational timescale of our scheme. Also, the thermal broadening of the atomic beam momentum distribution in this $\mu K$ temperature regime is small compared to the width of the central chaotic sea and thus would not contribute to the particle flux, hence keeping the efficiency of our scheme unaffected.

Another possible realization is provided by using a superconducting quantum interference device (SQUID) setup with Josephson junctions as in {\cite{Spiechowicz2014}} operating in the underdamped classical regime of temperatures $\sim$ 1K (in which damping, noise as well as quantum effects can be neglected safely) with a time dependent biharmonic external flux. Since this underdamped classical regime has already been realized experimentally, we believe that realizing our scheme using such setup is possible{\cite{PhysRevLett.94.247002,PhysRevB.37.1525,PhysRevLett.79.3046,Ruggiero1999}}. Finally, we note that our control scheme is not restricted to two lattice systems, but could be applied also to other Hamiltonian systems having mixed phase spaces and offering chaotic layers which can be systematically connected and disconnected.
%

\section{Conclusions} 
We provide a scheme offering the real-time control of directed currents in superimposed driven lattices. It can be straightforwardly implemented in ac-driven optical lattices and allows to design directed currents of cold thermal atomic ensembles which can be consecutively accelerated, slowed and reverted on demand. The mechanism underlying our scheme operates in phase space and depends only on large scale structures like the presence of different chaotic layers and cantori structures and should therefore be applicable more generally, e.g. as a design principle for targeted delivery of molecules or colloids in optical devices, or possibly on other vibrating substrates.

\begin{acknowledgments}
B.L. gratefully acknowledges funding by a
Marie Sk\l{}odowska-Curie Intra European Fellowship (G.A. no 654908)
within Horizon 2020. A.K.M thanks the Deutscher Akademischer Austauschdienst (DAAD) for funding through a doctoral research grant (Funding ID: 57129429).
\end{acknowledgments}

\bibliographystyle{apsrev}
\bibliography{mybib}

\end{document}